\documentclass[11pt,fleqn]{article}
\usepackage{amsmath,graphicx}

\newcommand{\be}{\beq\label}
\newcommand{\ee}{\eeq}

\usepackage{amsfonts,amssymb,cite}
\usepackage{epsf,graphicx}

\def\noi{\noindent}

\newcommand{\Title}[1]{\noi {{\Large\bf #1}}\\[1ex]}

\def\Aunames#1{\noi{\bf #1}}
\def\auth#1{${}^{#1}$}
\def\Addresses#1{\medskip\noi \protect
	\begin{description}\itemsep -3pt {\it #1} \end{description}}
\def\addr#1#2{\item[${}^{#1}$]{\it #2}}

\newcommand{\Abstract}[1]{\vskip 2mm \begin{center}
        \parbox{16.4cm}{\small\noi #1} \end{center}\medskip}
\newcommand{\PACS}[1]{\begin{center}{\small PACS: #1}\end{center}}

\def\email#1#2{\footnotetext[#1]{e-mail: #2}\addtocounter{footnote}{1}}

\def\nqq{\hspace*{-2em}}

\def\Jl#1#2{#1 {\bf #2},\ }

\def\ApJ#1 {\Jl{Astroph. J.}{#1}}
\def\CQG#1 {\Jl{Class. Quantum Grav.}{#1}}
\def\DAN#1 {\Jl{Dokl. AN SSSR}{#1}}
\def\GC#1 {\Jl{Grav. Cosmol.}{#1}}
\def\GRG#1 {\Jl{Gen. Rel. Grav.}{#1}}
\def\JETF#1 {\Jl{Zh. Eksp. Teor. Fiz.}{#1}}
\def\JETP#1 {\Jl{Sov. Phys. JETP}{#1}}
\def\JHEP#1 {\Jl{JHEP}{#1}}
\def\JMP#1 {\Jl{J. Math. Phys.}{#1}}
\def\NPB#1 {\Jl{Nucl. Phys. B}{#1}}
\def\NP#1 {\Jl{Nucl. Phys.}{#1}}
\def\PLA#1 {\Jl{Phys. Lett. A}{#1}}
\def\PLB#1 {\Jl{Phys. Lett. B}{#1}}
\def\PRD#1 {\Jl{Phys. Rev. D}{#1}}
\def\PRL#1 {\Jl{Phys. Rev. Lett.}{#1}}

\def\lal{&&\nqq {}}

\def\beq{\begin{equation}}
\def\eeq{\end{equation}}
\def\bear{\begin{eqnarray}}
\def\bearr{\begin{eqnarray} \lal}
\def\ear{\end{eqnarray}}
\def\earn{\nonumber \end{eqnarray}}

\begin{document}
\twocolumn[

\Title{Conservative relativistic algebrodynamics \\ induced  on an implicitly defined worldline }

\Aunames{Abdel Challa\auth{a,1}, Vladimir V. Kassandrov\auth{a,2} and Nina V. Markova\auth{b,3}}

\Addresses{
\addr a {Institute of Gravitation and Cosmology, Peoples' Friendship
	University of Russia (RUDN University), 6 Miklukho-Maklaya St., Moscow, 117198, Russian Federation}
\addr b {S.M. Nikolsky Mathematical Institute, Peoples' Friendship
	University of Russia, (RUDN University), 6 Miklukho-Maklaya St., Moscow, 117198, Russian Federation}
	}

\Abstract
  {In the framework of the Stueckelberg-Wheeler-Feynman concept of  a ``one-electron Universe'' we consider a worldline implicitly defined by a system of algebraic (precisely, polynomial) equations. Collection of pointlike ``particles'' of two kinds on the worldline (or its complex extension) is defined by the real (complex conjugate) roots of the polynomial system and detected then by an external inertial observer through the light cone connections. Then the observed collective dynamics of the particles' ensemble is, generally, subject to a number of Lorentz invariant conservation laws. Remarkably, this poperty follows from the Vieta's formulas for the roots of generating polynomial system.  At some discrete moments of the observer's proper time the mergings and subsequent transmutations of a pair of particles-roots take place simulating thus the processes of annihilation/creation of a particle/antiparticle pair}


] 
\email 1 {abdeltchalla@gmail.com}
\email 2 {vkassan@sci.pfu.edu.ru}
\email 3 {n.markova@mail.ru}

\section{``Unique worldline'' and \\ collective algebraic dynamics}
\label{sec1}

In the Special Theory of Relativity, any worldline canonically parameterized as $x_\mu = f_\mu (\tau)$ by a monotonic parameter $\tau$ describes the motion of a sole pointlike particle. The problem of interaction is actually the problem of mutual arrangement of different worldlines (H. Minkowski~\cite{Mink}) which, kinematically, can be quite arbitrary.  

In 1933 E.C.G. Stuekelberg~\cite{Stueck} examined worldlines of general type strictly prohibited in the canonical STR: such worldlines allow for segments with superliminar velocities and/or motions backwards in time. However, if one introduces another monotonic timelike parameter $t$ independent of the structure of the worldline itself, then at each $t$ one will detect a number of particles all moving forward in time. 

As for admissible superliminar values of velocities, they occure, as a rule, only in vicinity of critical points where a pair of particles merge and disappear (or arise). Such events can be naturally identified with physical processes of annihilation/creation of a particle/antiparticle pair. Note that just in this picture the famous Feynman's idea on a ``positron as an electron moving backwards in time'' had found its origin~\cite{Feynman1}.

Unfortunately, Stueckelberg's approach has a number of evident drawbacks. It leads to consideration of a 5-dimensional space with two timelike parameters and, in fact, to nonrelativistic kinematics with respect to the introduced ``true time'' $t$. It does not give the key to explanation of particle/antiparticle asymmetry. Moreover, one can see that at the points, say, of annihilation, ``matter disappears'' completely, so that none conservation laws can be valid. 

Nonetheless, the idea of Stueckelberg and his followers, J.A. Wheeler and R.P. Feynman~\cite{Feynman2}, looks quite attractive since it offers a way to construct the whole collective dynamics from solely simplest considerations purely kinematical in origin. To realize this idea, in~\cite{JPhys1} we had proposed to define a general worldline {\it implicitly}, that is, by means of a system of three algebraic equations
\be{implicit}
F_a (x_1,x_2,x_3,t) =0, ~~~~a=1,2,3. 
\ee
Then for any fixed value of the monotonically increasing timelike parameter $t$ one observes a number of pointlike particles located at $x_a (t)$ and determined by the roots of generating system (\ref{implicit}). In~\cite{Vestnik,JPhys2} we restricted by the functions $F_a$ {\it polynomial} in all $x_a$ as well as in $t$. Indeed, only in polynomial case the complete set of the roots of system (\ref{implicit}) can be easily found. Apart from this, just for polynomial equations different roots are bound by the set of {\it Vieta's formulas}, and this property leads directly to a set of {\it conservation laws} for the ensemble of particles-roots under consideration~\cite{Vestnik}. 

On the other hand, roots of a polynomial system of equations (with real-valued coefficients) are either real (R) or complex conjugate (C). The latters cannot be discarded as ``nonphysical'': they enter the Vieta's formulas and contribute,  as a result, to the conservation balance on equal right  with the real-valued roots. We conclude thus that the ensemble in question consists of two different kinds of particles, R- and C-, respectively.  

One can depict a conjugate pair of C-particles in the real 3D space marking equal real parts of their coordinates. Then one finds that a C-particle is located aside of the worldline's trajectory (oftenly, between isolated branches of the latter). 

In the course of time $(t)$ some two real roots (R-particles) approach one another (on one fixed branch of the trajectory) and, at some $t=t_0$, merge and convert then (at $t>t_0$) into a pair of complex conjugate roots, that is, a C-particle. The latter, after travelling for some time across the ``complex extension'' of the worldline, can reach another branch of the real trajectory and give there rise to a pair of R-particles scattering along the trajectory. Contrary to the Stueckelberg's scheme, the whole set of conservation laws holds in such annihilation/creation processes~\cite{Vestnik}.

It should be noted that on physical 3D space generating equations (\ref{implicit}) define in fact {\it non-relativistic} dynamics with quasi-Newtonian absolute time $t$. To pass to a {\it Lorentz invariant generalization} of algebraic dynamics, one can introduce an external ``observer'' moving along its own worldline and detecting the positions of particles on the generating worldline through the Lorentz invariant {\it light cone connections}. 

 Specifically, in~\cite{JPhys2} we considered a {\it canonically parameterized} worldline $x_\mu (\tau)$ and an observer {\it at rest} linked with points on the worldline by the {\it light cone equation}
 \be{cone1}
 (T-t(\tau))^2 -x(\tau)^2-y(\tau)^2-z(\tau)^2 =0,
 \ee
 $T$ being the {\it macroscopic} monotonically increasing proper time of the observer. Even in this simplest case, for {\it polynomial} parameterization (as well as for rationally parameterized worldlines, see~\cite{Rational}) the induced collective dynamics turns out to be {\it conservative}. That is, a full set of Lorentz invariant conservation laws holds for the collection of RC- particles-roots defined by (\ref{cone1}) as 
$\tau=\{\tau_k (T)\}  \rightarrow  x(T),y(T),z(T) = \{x_k (\tau_k), y_k (\tau_k),z_k(\tau_k)\}$. Note that for a {\it non-inertial} observer the conservation property is no longer valid. 
 
Remarkably, a number of unexpected effects has been revealed for {\it asymptotically great} values of the observer's time $T>>1$ if only $n>m$, ~$n$ and $m$ being the degrees of polynomials $t(\tau)$ and the oldest of $x_a(\tau)$, respectively.  Indeed, under these conditions near some critical value of $T=T_0$ one encounters the phenomenon of {\it pairing} of the roots while at greater values $T>>T_0$ the arised pairs manifest  the effect of {\it clusterization}, that is, formation of closely located (and generally large) groups of roots-particles~\cite{JPhys2,Thesis}. In the case of rationally parameterized worldlines~\cite{Rational} the emergent dynamics is still more diverse.   

Below in the paper, we combine the idea of implicit definition of a worldline represented by Eq. (\ref{implicit}) with the process of detection of the particles' dynamics on it through the light cone equation for an inertial observer.  For simplicity, we restrict by consideration of {\it plane} dynamics. In section 2, we briefly review the results of~\cite{JPhys1,Vestnik,JPhys2}: we pay attention to the origination of conservation property and present then an illustrative example of a polynomial 2D dynamics. Then, in section 3, the relativistic invariant dynamics is considered, in particular, for the same worldline as before. However,  particles on the worldline are now linked with an observer at rest by the light cone equation. In sectiion 4, a set of conservation laws is again obtained and represented in a manifestly relativistic invariant form. The asymptotic behaviour is also examined.

\section{Plane algebraic dynamics on an implicit polynomial worldline}

Consider a general system of {\it two} polynomial equations of orders $n$ and $m$, respectively:
\be{poly}
\begin{array}{ll}
F(x,y,t)=\sum_{p=0}^{p=n} \sum_{i,j,k}^{i+j+k=n} A_{ijk} x^i y^j t^k =0,\\
G(x,y,t)=\sum_{p=0}^{p=n} \sum_{i,j,k}^{i+j+k=m} B_{ijk} x^i y^j t^k =0,
\end{array}
\ee
where $A_{ijk}, B_{ijk}$ are the sets of arbitrary real coefficients (below, to simplify the computer algebra based analysis, we set them  to be integers).  

Let us assume that polynomials $F$ and $G$ do not contain a common factor and the leading coefficients are nonzero. Then for any value of the timelike parameter $t$ system (\ref{poly}) has exactly $n\times m$ roots, real valued or complex conjugate~\cite{Besu}. Thus, (\ref{poly}) induces a 2D algebraic dynamics of a set of R- and C- particles on the plane $(XY)$.

 The roots can be found via the procedure of {\it elimination} of one of the unknowns, say, $y$ from the two equations (\ref{poly}). In result, one comes to an equation for the so-called {\it resultant} of two 
polynomials $R(x,t)$ which itself is a general polynomial in $x$ and $t$ of the degree $N=n\times m$, 
\be{res}
R(x,t)=\sum_{p=0}^{p=N}\sum_{i,k}^{i+k=p} C_{ik} x^i t^k =0,
\ee     
$C_{ik}$ being the integers depending on $A_{ijk},B_{ijk}$; under the above formulated assumption, the leading coefficient $C_{N0}$ cannot turn to zero. Analogous procedure for elimination of $x$ obviosly leads to a dual equation for the resultant $\tilde R(y,t)$, 
\be{resd}
\tilde R(y,t)=\sum_{p=0}^{p=N} \sum_{i,k}^{i+k=p} \tilde C_{ik} y^i t^k =0, 
\ee
and, properly combining the N roots of (\ref{res}) with $N$ roots of (\ref{resd}), one obtains, for any $t$,  a complete set of $N$ solutions $\{x_k(t),y_k(t)\},~k=1,2,\cdots N$ to the original system of equations (\ref{poly}). 

The monoms in (\ref{res}) and (\ref{resd}) can be equivalently rearranged in the following way:
\be{res2}
\begin{array}{ll}
R(x,t)=a_0 x^N + a_1(t) x^{N-1} + \cdots +a_N (t) =0, \\
\tilde R(y,t)=b_0 y^N + b_1(t) y^{N-1} + \cdots  +b_N (t) =0, 
\end{array}
\ee
where $a_0,b_0$ are constants, while $\{a_k(t), b_k(t)\},~k=1,2\cdots N$ are {\bf polynomials of degree (no more than) $k$ in $t$}.

Now, from the first two (modified, see, e.g.,\cite{Newtonident}) {\it Vieta's formulas} linking  $N$ roots of the Eqs. (\ref{res2}) it follows:
\be{Viet1}
\begin{array}{ll}
x_1+x_2+\cdots x_N = -a_1(t)/a_0, \\ 
y_1+y_2+\cdots y_N = -b_1(t)/b_0,
\end{array}
\ee
and
\be{Viet2}
\begin{array}{ll}
x_1^2+x_2^2 + \cdots +x_N^2 = -2 a_2(t)/a_0 + (a_1(t)/a_0)^2, \\ 
 y_1^2+y_2^2 + \cdots +y_N^2 = -2 b_2(t)/b_0 + (b_1(t)/b_0)^2.
 \end{array} 
\ee
Note that complex conjugate roots enter in pairs so that l.h.s. in (\ref{Viet1},\ref{Viet2}) are real-valued. The r.h.s. {\bf are linear or quadratic} in $t$, respectively. Differentiating these by $t$, one obtains
\be{momxy}
\begin{array}{ll}
\dot x_1+\dot x_2 + \cdots +\dot x_N = constant:=P_x, \\
 \dot y_1+\dot y_2 + \cdots +\dot y_N = constant:=P_y, \\
 \end{array}
 \ee 
  and
 \be{virialxy}
 \begin{array}{ll}
(\dot x_1^2 +x_1 \ddot x_1)+ \cdots +(\dot x_N^2 +x_N\ddot x_N )= constant, \\ 
(\dot y_1^2 +y_1 \ddot y_1)+ \cdots +(\dot y_N^2+y_N \ddot y_N) = constant.
 \end{array} 
 \ee
 Whether one assumes that masses of all R-particles are equal (so that any C-particle possesses twice greater mass), relations (\ref{momxy}) represent the {\it law of conservation} of total momentum of the set of RC- particles (for  the projections $P_x$ and $P_y$, respectively). As to relations (\ref{virialxy}), after summing these up one comes to a conservation law for a SO(2) scalar -- an analogue of the total energy of the RC ensemble which is closely related to the {\it virial} in classical mechanics (for detail, see e.g.~\cite{Virial}):
\be{virial}
(\dot r_1^2 +\vec r_1 \ddot {\vec r}_1)+ \cdots +(\dot r_N^2 +\vec r_N\ddot {\vec r}_N )= constant:=E,   
\ee            
where $\{\vec r_k, \dot {\vec r}_k, \ddot {\vec r}_k\} $ are the radius-vector, velocity and acceleration of the $k$-th particle, respectively.

Finally, numerous calculations with different generating polynomials (\ref{poly}) allow to conclude that {\it conservation of the total angular momentum} $\vec M$ holds for any irreducible pair of them (in the 2D case of plane dynamics vector $\vec M$ has only one nonzero component),  
\be{angulmom}
M_z =\sum_i (x_i \dot y_i - y_i \dot x_i).
\ee
Special procedure which allows to analytically compute $M_z$ for any given pair of generating polynomials $F$ and $G$ has been elaborated in~\cite{Vestnik,JPhys2}; it will be examplified below. 

Let us briefly illustrate the above scheme on a randomly selected~\footnote{To simplify the posterior computations in Sect.3, we only take the second polynomial to be linear in t} nondegenerate polynomial system 
\be{exmpl}
\begin{array}{lll}
 F:=x^3+2x^2y +2t^3-t-3 =0, \\
 G:=-2x^3+y^3 + 6txy-3x^2y \\
 +(x+y)t+y+2=0,
\end{array}
\ee
at any $t$ possessing $3\times 3 = 9$ roots some of which correspond to R- particles while others, complex conjugate, -- to C-particles (with a twice greater mass). Resolving the second equation w.r.t. $t$ and substituting the result into the first one we obtain the {\it equation of trajectory} of R-particles which turns out to contain three isolated branches (Fig.1). 

\begin{center}
\begin{figure}[ht]
\centering
\includegraphics[angle=0,scale=0.7]{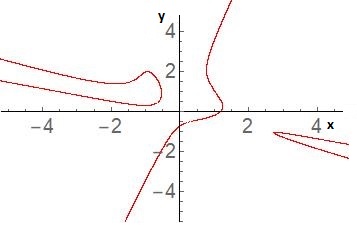}
\vskip 1.7 cm
\caption{\small Three isolated branches of a ``unique worldline'' trajectory for R-particles following from (\ref{exmpl})}
\label{pic2}
\end{figure}
\end{center}   

By finding then corresponding resultants to eliminate $y$ or $x$, respectively, we obtain
\be{resexmpl}
\begin{array}{ll}
R(x,t) = 5x^9 +24 t x^8-(4t-4)x^7+\cdots =0,\\
R(y,t)= 5y^9-12ty^8+(9t+11)y^7+\cdots =0.
\end{array}
\ee
Seeking now for values of $t$ at which both resultants have multiple roots, one gets four such (real) moments, namely $t\approx -0.9576, 0.2565, 1.2896, 1.2995$. At each of these moments one encounters an event of annihilation or creation of a pair of R-particles; maximal number (five) of R-particles exists within the time interval $1.2896<t<1.2995$. 

Now, making use of the (modified) Vieta formulas, one gets
 \be{vietexmpl}
 \begin{array}{lll}
 \sum_{i=1}^9 x_i = -24t/5, ~~~~\sum_{i=1}^9 y_i = 12t/5, \\
  \sum_{i=1}^9 x_i^2 = (24t/5)^2 +2(4t-4)/5, \\ 
  \sum_{i=1}^9 y_i^2 = (12t/5)^2-2(9t+11)/5.
 \end{array}
 \ee  
After differentiation (\ref{vietexmpl}) by $t$, we come to a conserved value of the total momentum, 
\be{momexmpl}
P_x = \sum_{i=1}^9 \dot x_i = -24/5 ~~P_y=\sum_{i=1}^9 \dot y_i = 12/5 \\
\ee               
and to a constant value of the SO(2) scalar, the analogue of total energy, 
\be{virexmpl}
E:=\sum_{i=1}^9 (\dot x_i^2+\dot y_i^2) +(x_i \ddot x_i+y_i \ddot y_i)= 144/5.  
\ee
As for a value of the total angular momentum, to find it we can operate as follows. Taking total derivatives by $t$ of the initial system (\ref{poly}) and resolving the arising linear system w.r.t. velocities $\dot x,\dot y$, we express then the $Z$-component of angular moment (\ref{angulmom}) through $x,y,t$, factorize the relation $\mu_z - f(x,y,t)=0$ and take the numerator to obtain a new polynomial equation of the form $H(\mu_z,x,y,t)=0$. Eliminating then (by taking subsequent resultants) $x$ and $y$ from the system formed by this equation together with two initial ones (\ref{poly}), we get an equation of the form~\footnote{A redundant factor could arise in the procedure which should be discarded}
\be{preservang}
Q_0 \mu_z^9 +Q_1 \mu_z^8 +\cdots + Q_9(t) =0,
 \ee    
where the first two coefficients $Q_0$ and $Q_1$ always happen to be constant by $t$. Therefore, making use of the first two Vieta's formulas, for the value of total angular momentum $M_z$ one obtains:
\be{sumang}
M_z =\sum_{k=1}^9 \mu_z = -Q_1/Q_0 =constant.
\ee    
However, for the particular example of the system 
(\ref{exmpl}) $Q_1=0$, so that in this case $M_z$  is precisely null. 

For interpretation of the dynamical pattern one should examine the {\it asymptotic} regime at  great values of $t$. To do this, let us divide Eqs.(\ref{exmpl}) by $t^3$ and neglect the terms of order $1/t$ or smaller. Then for the ratios $X:=x/t,~Y:=y/t$ one obtains a system of two equations with constant coefficients, 
\be{asympt}
\begin{array}{ll}
X^3+2 X^2 Y+2=0,\\ 
-2X^3+Y^3+6XY-3X^2Y=0,
\end{array}
\ee
which has 3 real roots and 3 pairs of complex conjugate of those. This means that asymptotically, at great negative $t$, 3R- and 3C-particles approach each other moving uniformly towards the origin (some of them with superluminar velocities) while for great positive values of $t$ the same number of them uniformly scatter to infinity. Asymptotic picture ``at the future'' as a whole reproduces that ``at the past'' though at small values of $t$ one encounters actually a cascade of mutual transmutations of the two kinds of pointlike particles. Certainly, such a picture is typical for any polynomial worldline defined by Eqs.(\ref{poly}).

The above considered algebraic dynamics is completely non-relativistic. The time $t$ monotonically increase, being absolute, universal and actually analogous to the Newtonian time. In order to pass to relativistic generalization we shall input below an external ``observer'' detecting the particles on a worldline  through the ``signals'' continiously propagating with the speed of light  from the points on the worldline to the observer. 

\section{Light-detected collective algebraic dynamics on an implicitly defined worldline}

Consider an observer {\it at rest} at the origin of (XY) plane receiving the light-like signals from the points on a worldline implicitly defined by a polynomial system of equations of the form (\ref{poly}). 
Then we have to supplement these equations by the equation of light cone 
\be{cone}
S:=(T-t)^2 - x^2 - y^2 =0, 
\ee
 $T$ being the (monotonically increasing) {\it proper time} of the observer. For any $T$, one has to find a complete set of roots $\{x,y,t\}$ of the system (\ref{poly}) and (\ref{cone}). Remarkably, the trajectory of R-particles following from (\ref{poly}) will remain the same, only the law of their motion w.r.t. $T$ will change.  
 
Even the number of particles-roots observed now for any $T$ will be twice greater than it is for any $t$ according to the system (\ref{poly}). Moreover, in the course of time $T$ individual times $\{t_k\}$ can decrease and/or correspond to the ``advanced'' signals, signals ``from future''. Together with locations of C-particles at a complex extension of the (XY) plane, the times $\{t_k\}$ can be even complex-valued. Actually, the sole physical time in the scheme is the ``macroscopic'' time of the observer $T$. 

Since the equation of light cone (\ref{cone}) is a nondegenerate polynomial equation of second order, the collective dynamics defined by the system (\ref{poly},\ref{cone}) will be again {\bf conservative}. The proof can be obtained by a direct generalization of the procedure described in Sect.2 and is described in detail in~\cite{JPhys2,Thesis}. 

Let us consider the same physical RC-system as it is defined by the equations (\ref{exmpl}) but w.r.t. the light-detecting ``observer'' represented by the light cone equation (\ref{cone}). We resolve the second equation in (\ref{exmpl}) w.r.t. $t$ and substitute it into the first equation (\ref{exmpl}) and (\ref{cone}). Eliminating then one of the unknowns, say, $x$ by taking corresponding resultant and repeating the procedure for the other unknown $y$, we get (compare with ({\ref{resexmpl})):
\be{resT}
\begin{array}{llll}
r(y) =5430955y^{18} -(19067340 T - 5278920)y^{17} + \\ (20644236 T^2   
-9522840 T - 1074680)y^{16}+\cdots =0,\\
r(x) = 5430955x^{18}+(11335920 T + 4714960)x^{17} \\                                          -(13323096 T^2  - 4805150 T + 25495)x^{16}+\cdots =0.                                    
\end{array}
\ee

Making now use of the modified Vieta's formulas and reproducing the procedure exposed in Sect.2, we obtain the following conservation laws:
\be{conservLorentz}
\begin{array}{ll}
P_x =3813468/1086191,P_y=-2267184/1086191, \\ 
E=16501798925784/1179810888481.
\end{array}
\ee

Another peculiarity of the observer-dependent scheme for algebraic dynamics is the emergence of two new conserved quantities -- the temporal component of the conserved total 4-momentum of the RC-ensemble~\footnote{In the considered example of plane 2D dynamics one of the spacial components of the momentum 4-vector is identically zero}, and corresponding second order quantity:
\be{temporl}
P_t:=\sum_i \dot t_i, ~~~V_t: = \sum_i \dot t_i^2 + t_i \ddot t_i. 
\ee 
To determine $P_t$ and $V_t$, one should eliminate both unknowns $x$ and $y$ to obtain the implicit dependence
\be{restemp}
\begin{array}{ll}
r(t) =1086191t^{18} - (219870 T-15598)t^{17} + \\
(2198187T^2+779024T -1142373)t^{16}+\cdots =0 ,
\end{array}
\ee
and again make use of the modified Vieta's formulas and their derivatives. In particular, for the example in question (\ref{exmpl},\ref{cone}) one gets
\be{temprPV}
\begin{array}{ll}
P_t =2190870/1986191, \\ V_t =24609485466/1179810888481. 
\end{array}
\ee
 Now we can compute the {\it 4-scalars} corresponding to the 4-momentum and to the 
 analogue of relativistic energy of the system, namely, 
 \be{4scalars}
 M^2:= P_x^2 +P_y^2-P_t^2, ~~~W:=V_x+V_y -V_t. 
 \ee
The first quantity should be identified with the total rest mass of the RC-system. 
Numerical values of the above 4-scalars in the considered example turn out to be
\be{masssq}
M^2 = 13701780/1086191,~W=151697/1086191. 
\ee
Finally, the nonzero Z-component $M_z$ of the total angular momentum vector is also conserved and, in the particular example, equal to 
\be{angmom}
M_z=-472656/1086191.
\ee

\section{Concluding remarks}

To begin with, we note that the 3D generalization of the above considered plane (2D) dynamics can cause only technical problems (see~\cite{Vestnik,Thesis}) and none of fundamental nature. 
We have also seen that the ``subjective'' algebraic dynamics detected on an implicitly defined  worldline by an external ``observer'' through the light connections essentially differs from the ``objective'' one determined by the equations of the worldline itself (\ref{poly}). Nonetheless, 
in both situations one encounters a set of pointlike ``particles'' of two kinds. 

 The first kind, R-particles, is determined by real-valued roots of the generating algebraic system of equations while the second, C-particles, correspond to complex conjugate roots and live aside of the worldline trajectory. For the whole set of RC-particles the induced ``polynomial'' dynamics turns out to be conservative both in ``objective'' and ``subjective'' cases. However, in the latter case the observer should be inertial. 
 
Remarkably, the set of relativistic fields generated by any of the particles (for detail see, e.g.,~\cite{IJGMMP}) do not explicitly contribute to the structure of conservation relations. This allows  to formulate conservation laws in a closed and manifestly Lorentz invariant form what is in  principle impossible in the framework of the canonical STR. 

Specifically, considering the masses of all particles be equal and setting these be unit, we obtain the law of conservation of total 4-momentum in the form 
\be{4momconserv}
\sum_i \dot x_\mu^{(i)} = constant, 
\ee
where $(\cdot)$ denotes differentiation w.r.t. the unique proper time of the observer. Corresponding 4-scalar, total rest energy (rest mass), is 
\be{restenergy}
M^2 = \sum_i \dot x_\mu^{(i)}\dot x_\nu^{(i)}\eta^{\mu\nu} = constant,  
\ee
where tensor $\eta^{\mu\nu}$ represents the Minkowski metric.  

The other conserved 4-scalar related to the Vieta's formulas of second order has the form
\be{4scalar2}
W=\sum_i (\dot x_\mu^{(i)}\dot x_\nu^{(i)}+\ddot x_\mu^{(i)} x_\nu^{(i)})\eta^{\mu\nu} = constant,  
\ee
and could be identified as the analogue of total energy. It differs from (\ref{restenergy}) by the second term which is proportional to the particles' accelerations and could substitute the energy of radiation. 

Recall that under consideration of a polynomial {\it explicitly parameterized} worldline~\cite{JPhys2,Thesis} both invariants coincide in value, $W\equiv M=2p\ge 0$, are positive definite and, moreover, ``self-quantized'' being equal to twice the degree $p$ of the polynomial parameterizing the temporal coordinate $x_0 (\tau)$. Unfortunately, for implicitly defined worldlines this property does not hold. 

Finally, as we have already presented, numerical experiments ``highly likely'' demonstrate that all 6 components of the total angular momentum skew tensor
\be{angtensor}
M_{[\mu\nu]}:= \sum_i x_\mu^{(i)} \dot x_\nu^{(i)} -x_\nu^{(i)} \dot x_\mu^{(i)} = constant. 
\ee            
are also preserved for any polynomial implicit worldline (\ref{poly}). 

The asymptotic dynamic behaviour, now w.r.t. the invariant proper time $T$, can be obtained in the same manner as in the previous scheme of internal non-relativistic dynamics. Particularly, for the system in question (\ref{poly},\ref{cone}) for the ratios $X:=x/T,~Y:=y/T,~S:=t/T$ one gets a system of 3 equations which has 6 real roots and 6 pairs of complex conjugate ones. Thus, we come again to a picture of ``shrinking and then expanding Universe'' in which the final stage  completely reproduces the starting one. Unfortunately, there are no indications for coupling or formation of clusters at late stage of evolution, contrary to the emergence of these phenomena in the corresponding model with an {\it explicitly parameterized} polinomial~\cite{JPhys2} or rational~\cite{Rational}  worldline.

The authors are grateful to I.Sh. Khasanov for his help in the computer algebra applications.  
The publication has been prepared with the support of the ``RUDN University Program 5-100''.

\end{document}